\documentclass[useAMS,usenatbib]{mn2e}
\usepackage{amsmath}
\usepackage{amssymb}
\usepackage{multicol}
\usepackage{rotating}

\title[M. E. Bell et al]
{A search for variable and transient radio sources in the extended \emph{Chandra} Deep Field South at 5.5 GHz}
\author[M.E.Bell et al]{M. E. Bell$^{1,2,3}$\thanks{E-mail:
martin.bell@csiro.au (MEB)}, M. T. Huynh$^{4}$, P. Hancock$^{2,3,5}$, Tara Murphy$^{2,3}$, B. M. Gaensler$^{2,3}$, 
\newauthor D. Burlon$^{2,3}$, C. Trott$^{2,5}$, K. Bannister$^{1,2}$ \\
$^{1}$CSIRO Astronomy and Space Science, PO Box 76, Epping NSW 1710, Australia\\
$^{2}$ARC Centre of Excellence for All-sky Astrophysics (CAASTRO)\\
$^{3}$Sydney Institute for Astronomy (SIfA), School of Physics, The University of Sydney, NSW 2006, Australia\\
$^{4}$International Centre for Radio Astronomy Research, M468, University of Western Australia, Crawley, WA 6009, Australia\\
$^{5}$International Centre for Radio Astronomy Research, Curtin University, Bentley, WA 6845, Australia}

\begin{document}

\date{Accepted xxxx December xx. Received xxxx December xx; in original form xxxx October xx}

\pagerange{\pageref{firstpage}--\pageref{lastpage}} \pubyear{2002}

\maketitle

\label{firstpage}
\begin{abstract}
We present a three epoch survey for transient and variables in the extended \emph{Chandra} Deep Field South at 5.5 GHz with the Australia Telescope Compact Array. A region covering $\sim$0.3~deg$^{2}$ was observed on timescales of 2.5 months and 2.5 years and typical sensitivities 12.1 $-$ 17.1~$\mu$Jy beam$^{-1}$ (1$\sigma$) were achieved. This survey represents the deepest search for transient and variable radio sources at 5.5 GHz. In total 124 sources were detected above the 5.5$\sigma$ level. One highly variable radio source was found with $\Delta S  > 50\%$ implying a surface density of $\sim$3~deg$^{-2}$. A further three radio sources were found with lower levels of variability equating to a surface density of $\sim$13~deg$^{-2}$ above a detection threshold of 82.3~$\mu$Jy. All of the variable sources have inverted radio spectra (between 1.4 and 5.5 GHz) and are associated with active galactic nuclei. We conclude that these variables are young gigahertz peaked-spectrum sources with active and self-absorbed radio jets. We explore the variability completeness of this sample and conclude that the fairly low levels of variability would only be detectable in 3$-$25\% of all sources within the field. No radio transients were detected in this survey and we place an upper limit on the surface density of transient events $ < 7.5$~deg$^{-2}$ above a detection threshold of 68.8~$\mu$Jy. 
\end{abstract}


\section{Introduction}
There is a vast amount of parameter space available to search for transient and variable radio sources. The physical mechanisms driving dynamic radio emission remain poorly constrained as a function of frequency and flux density. This is especially true at higher frequencies (e.g. 5 GHz and higher) where the field of view (and survey speed) of radio interferometers decreases. At a range of survey cadences and at frequencies around 1.4 GHz, the dynamic sky has been well characterised above $\sim$ one mJy (e.g. \citealt{Bower_2010}; \citealt{croft_2010}; \citealt{Bower_2011}; \citealt{croft_2011}; \citealt{croft_2013}). At 5~GHz and below one~mJy however, this is not the case.   

At 1.4 GHz one of the benchmark surveys in exploring the sub-mJy variable radio sky is that of \cite{Carilli}. In this survey, five archival repeat VLA observations of the Lockman Hole were examined for variable radio sources on timescales of 17 months and 19 days. Of the sources detected between $50 - 100~\mu$Jy, 2\% were found to be variable. The surface density of variable sources above $100\mu$Jy was found to be $\sim 18$~deg$^{-2}$. On the shorter timescales of one day to three months, \cite{Mooley_2013} examined archival observations of the extended \emph{Chandra} Deep Field South (eCDFS; \citealt{eCDFS}).  At 1.4 GHz a small fraction (1\%) of sources were found to vary significantly above 40~$\mu$Jy.  

At $\sim$5 GHz, the deepest variability survey to date is that of \cite{Becker}, which probed a flux density greater than one~mJy, an order of magnitude higher in flux than that of the Carilli et al. (2003) survey. The \cite{Becker} survey showed that the surface density of variable sources towards the Galactic Plane above one mJy is $\sim$1.6 deg$^{-2}$ on timescales of days to years. In comparison Gregory \& Taylor (1986) showed that a much stricter limit of $\sim 10^{-3}$ deg$^{-2}$ is found above 70 mJy on timescales of days and years. To date there has been no blind survey at 5~GHz that has studied the abundance of long duration variable radio sources below 1 mJy. 

A few blind radio surveys have searched for both transients and variables at 5 GHz on long timescales. For example, \cite{Bower_2007} (also see \citealt{Frail_2012}) searched archival data (at 5 and 8 GHz) of a single calibration field observed over 22 years with the Very Large Array (VLA). Upper limits were placed on the surface density of long duration transients. 
\cite{Ofek_2011} searched for both transients and variables using the VLA at 5 GHz. One possible radio transient was reported and 2\% of the source sample ($>0.5$~mJy) were found to vary on a timescale of two years. 

Here we present a new set of observations aimed at probing the time-domain properties of the radio sky at 5~GHz below 1~mJy.
The eCDFS is a $\sim$0.3~deg$^{2}$ region  favoured for its low optical and HI extinction. A plethora of multi-wavelength data from both space and ground based facilities are available for this region (e.g. Grogin et al. 2011, Xue et al. 2011, Koekemoer et al. 2011, Skelton et al. 2014). At 1.4~GHz, a 3.6~deg$^{2}$ region encompassing the eCDFS has been observed by the Australian Large Array Survey (ATLAS; \citealt{ATLAS}) team using the Australia Telescope Compact Array (ATCA). At both 1.4~GHz and 5~GHz, observations of the eCDFS have also been made using the VLA (see \citealt{VLA_CDFS}). In late 2009, \cite{CDFS_paper} obtained observations of the eCDFS at 5.5 and 9~GHz ($\sim 144$ hours). 

For the survey presented in this paper we obtained a further two epochs of observations ($\sim$100 hours) of the eCDFS at 5.5~GHz (designed to match those of \citealt{CDFS_paper}). 
The motivation for this survey was to search a bespoke region of parameter space (i.e. 5 GHz and $<$1~mJy) for signatures of dynamic radio emission on long timescales: months to years. It was also to survey a region which had extensive multi-wavelength coverage, thus to negate the need for potentially lengthly follow-up observations. The multi-wavelength data allow for accurate source identification, which has impeded the physical interpretation of discoveries in previous surveys. 

The outline of this paper is as follows. In sections 2 and~3 we describe the observations and data reduction procedure, respectively. In section 4 we describe the transient and variable search methodology. Sections 5 and 6 describe the source counts and characteristics of the variable and transient sources detected, respectively. In section 7 we discuss the physical interpretation of the results and we compare our work with that reported in the literature.
\section{Observations}
For the analysis presented in this paper we used archival data of the eCDFS field (PI: M. Huynh, project code C2028, see \citealt{CDFS_paper}) and new observations, both of which were obtained with the ATCA. We will refer to the archival data as epoch~1 henceforth. These archival observations were conducted between 2009~August~12 and 2010~January~16 and we include all data, regardless of the date collected, into this epoch. In total 115 hours hours of data were collected and the observations were typically organised into 12 hour blocks. Approximately 80\% of the observations were obtained in 2010~January. The final day of observations for epoch~1 were obtained on 2010~January~16 and we take this as the reference time for examining the variability. 

We obtained two further epochs of data for this survey (project code C2670). In Table \ref{obs_table} the dates, integration times and array configurations of the observations are summarised. The observations for epoch~2 were performed from 2012~May~31 and 2012~June~03 . The observations for epoch~3 were obtained between 2012~August~14 to 2012~August~19. In total we sampled timescales between epoch~1 and epoch~2 of 869 days (or 2 years, 4 months and 18 days) and between epoch-2 and epoch-3 of 77 days. This was calculated between the final dates of the observations for each epoch.  

We adopted the same observing strategy as \cite{CDFS_paper} to reproduce, as best as possible, the observations of epoch~1. We used 42 pointings in a hexagonal layout to mosaic a field of view $\sim30^{\prime} \times 30^{\prime}$. The 42 pointings were spaced such that at 5.5~GHz the map was sampled to uniform sensitivity. We observed each of the 42 pointings consecutively (interleaved with calibration observations) over the course of the observing block (as was done by \citealt{CDFS_paper}). For epochs 2 and 3 the observations were organised into five $\sim$10~hour blocks and executed intermittently over the dates specified above. We obtained approximately 48 hours of observations on epochs 2 and 3, which resulted in a $\sim \sqrt{2}$ decrease in sensitivity for these epochs, when compared with epoch~1.   
 
All of the images have been restored with a restoring beam of  $5.0^{\prime \prime} \times 2.0^{\prime \prime}$ with beam parallactic angle (BPA) 1.4 degrees.  We used the full available continuum bandwidth of 2048~MHz centred at 5.5 GHz using the Compact Array Broadband Backend \citep{CABB}. The observations were centred at RA~$3^{h}32^{m}2^{s}$ and Dec~$-27^{\circ}28^{\prime}34^{\prime\prime}$ (J2000). 

 \begin{table*}
\centering
\caption{A summary of observations used for this survey. Epoch~1 was obtained by Huynh et al. (2012). In total we have probed timescales of 869 days (between epochs 1 and 2) and 77 days (between epochs 2 to 3). The $N_{det}$ column gives the number of blind source detections in each image. The RMS column gives the typical noise calculated close to the centre of the image.}
\begin{tabular}{|c|c|c|c|c|c|c|}
\hline
\multicolumn{1}{c}{Epoch} & \multicolumn{1}{c}{Date} & \multicolumn{1}{c}{Total integration time} & Array & RMS & $N_{Det}$  \\
 &  &  (hours) & & ($\mu$Jy beam$^{-1}$) \\
\hline
1 & 2009 Jun 08 $-$ 2010 Jan 16& 115 & 6A, 6D & 12.1 & 115 \\
2 & 2012 May 31 $-$ 2012 Jun 03  & 48 & 6D & 17.1 & 82 \\
3 & 2012 Aug 14 $-$ 2012 Aug 19  & 48 & 6A & 16.2 & 91 \\
\hline
\label{obs_table}
\end{tabular}
\end{table*}
 
 \section{Data reduction}
There have been some improvements to the Multichannel Image Reconstruction, Image Analysis and Display (MIRIAD; \citealt{MIRIAD}) software package and the wide-band imaging steps used to reduce the data since the analysis of Huynh et al. (2012).  We follow the data reduction methodology described by Huynh et al. (in prep) but summarise it here briefly.  Epoch~1 from Huynh et al. (2012) was re-processed to be consistent with the latter epochs. Amplitude, phase and bandpass calibration solutions were derived from intermittent observations of PKS B1934-638 (amplitude and bandpass calibration) and PKS J0348-2749  (phase calibration). 

\begin{figure*}
\centering
\includegraphics[scale=0.80]{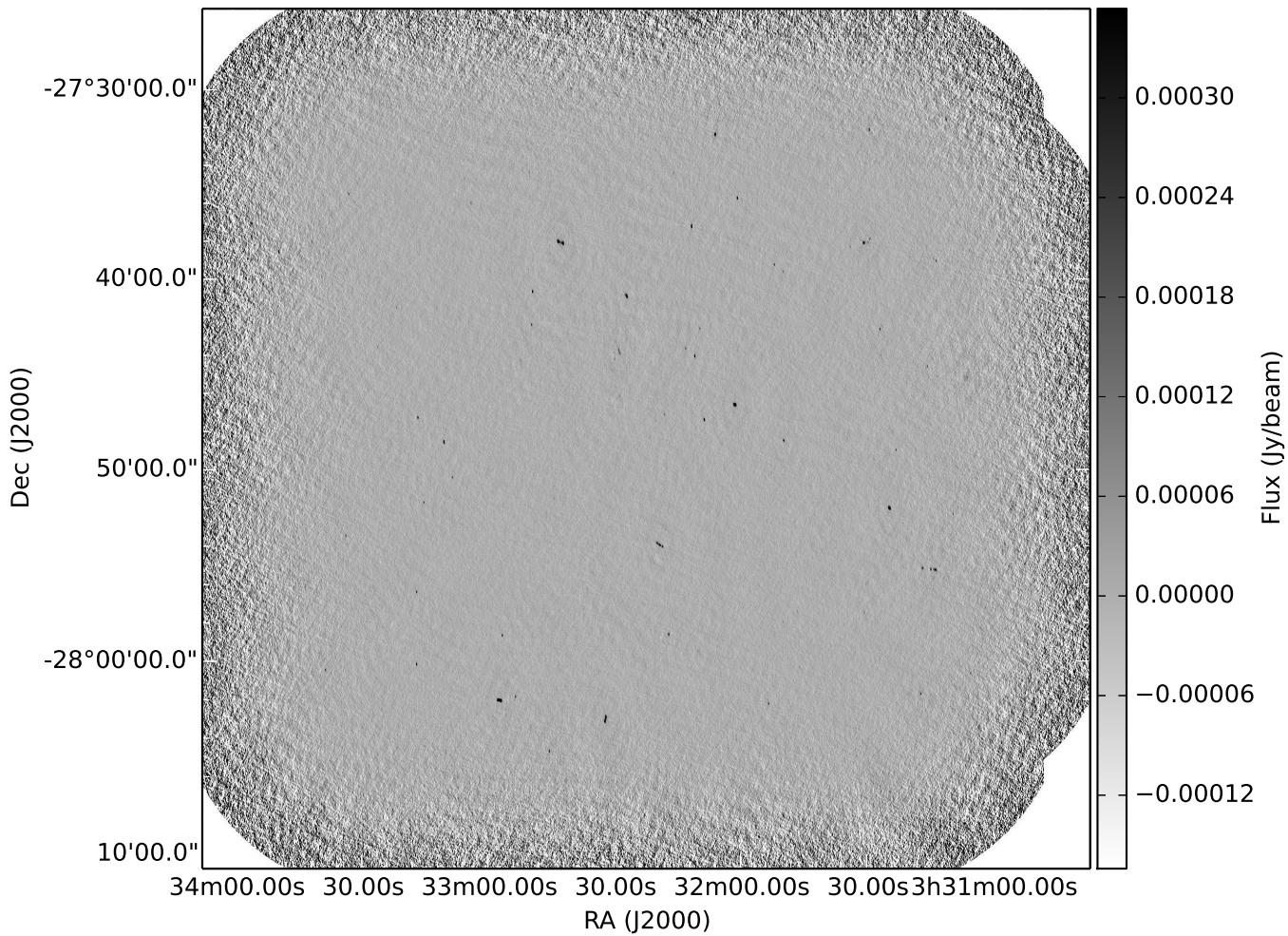}
\caption{An image of epoch~2. A total of $\sim$0.3~deg$^{2}$ was imaged using 42 separate pointings. In the central region of the map an r.m.s sensitivity of 17.1~$\mu$Jy beam$^{-1}$ was achieved. A total of 48 hours of data were used to make this image and 82 sources were detected. See Table \ref{obs_table} for details of other epochs. All images have been restored with a restoring beam of $5.0^{\prime \prime} \times 2.0^{\prime \prime}$ with beam parallactic angle (BPA) 1.4 degrees.}
\label{CDFS_image}
\end{figure*}

For each epoch of observations, each of the 42 pointings was deconvolved, CLEANed, and self-calibrated separately. The final image for each epoch was produced from a linear mosaic of the separate pointings. In Figure \ref{CDFS_image} we show an image of epoch~2. In epochs 1, 2 and 3 we achieved noise levels of 12.1, 17.1 and 16.2 $\mu$Jy beam$^{-1}$ respectively, close to the centre of the image (see also Table~\ref{obs_table}). 

\begin{figure}
\centering
\includegraphics[scale=0.60]{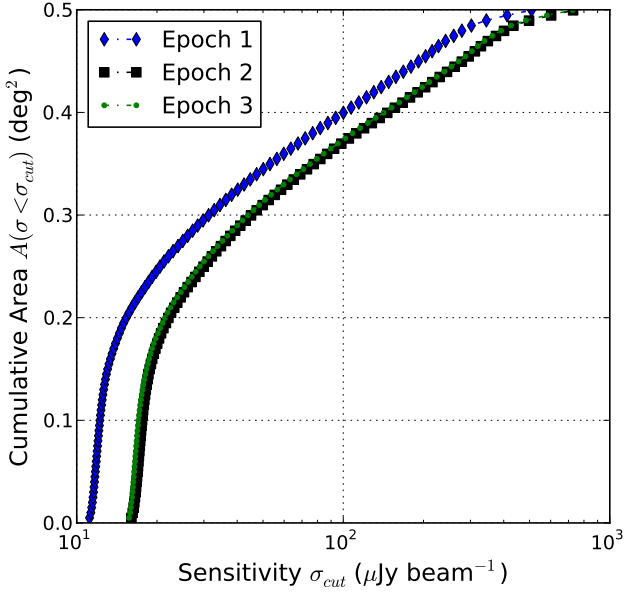}
\caption{Sensitivity $\sigma_{cut}$ ($\mu$Jy beam$^{-1}$) versus cumulative area $A$ (deg$^{2}$) as a function of epoch. The cumulative area gives the amount of sky we have observed with sensitivity $\sigma < \sigma_{cut}$. For example, in epoch~1 approximately 0.4~deg$^{2}$ is observed with sensitivity better than $100 \mu$Jy.}
\label{area_vs_sens}
\end{figure}

Figure \ref{area_vs_sens} shows the noise profile of the three epochs as a function of sky area. For each epoch we generated an r.m.s map, where each map pixel represented the 1$\sigma$ r.m.s sensitivity in that region. For each epoch we summed up the cumulative area (deg$^{2}$) and plotted this as a function of sensitivity. Figure \ref{area_vs_sens} shows that because the epochs are made of linear mosaics the noise profile is relatively flat out to $\sim$0.2~deg$^{2}$. Table \ref{area_vs_sens} shows the mean sensitivity over five cumulative sky area bins for epoch~1. These values are used in subsequent analysis to define the upper limit on flux density and area that this survey probed. 

\begin{table}
\centering
\caption{The mean sensitivity of epoch~1 calculated over five discrete sky area bins.}
\begin{tabular}{|c|c|}
\hline
Cumulative Area  & Mean Sensitivity   \\
 (deg$^{2}$) &  ($\mu$Jy beam$^{-1}$)    \\
\hline
0.1 & 11.7   \\
0.2 & 12.5   \\
0.3 & 15.5   \\
0.4 & 26.0   \\
0.5 & 63.0   \\
\hline
\label{area_sens_table}
\end{tabular}
\end{table}

\section{Transient and variable search techniques}
An automated pipeline developed as part of the Australian Square Kilometre Array Pathfinder (ASKAP; \citealt{ASKAP}) Variable and Slow Transients (VAST) project (\citealt{VASTPaper}) was used to search the observations for transient and variable radio sources. We refer the reader to \cite{Bell_2014} for a full description of the transient and variable search methodology. To summarise briefly, we located sources within the images using the {\sc aegean} source finder (see \citealt{Paul_AEGEAN}). We used a source extraction level of 5.5$\sigma$ to search for sources. A forced measurement algorithm was used to produce complete light-curves for all sources detected in one image greater than 5.5$\sigma$. For all the light-curves (constructed from either blind detections or blind detections plus forced measurements) we then tested for significant variability using the following statistics.  

Firstly, we tested for variability by calculating the $\chi^{2}$ probability that the source remained constant. The mean of the data points were used as a model to test if the light-curve was invariant. For each source we calculated $\chi^{2}_{lc}$ where

\begin{equation}
\chi_{lc}^{2} = \sum_{i=1}^{n} \frac{(S_{i} - \tilde{S})^{2}}{\sigma_{i}^{2}}.
\label{chisquared}
\end{equation}

\noindent $S_{i}$ is the i$th$ flux density measurement with variance $\sigma^{2}_{i}$. $n$ is the total number of epochs. The weighted mean flux density, $\tilde{S}$, is defined as 

\begin{equation}
\tilde{S} = \sum_{i=1}^{n} \left( \frac{ S_{i}}{\sigma_{i}^{2}} \right) /  \sum_{i=1}^{n} \left( \frac{1}{\sigma_{i}^{2}} \right).
\label{w_mean}
\end{equation}

The values of $\chi^{2}_{lc}$ are expected to follow the theoretical distribution $\chi^{2}_{T}$ (for $n-1$ degrees of freedom), assuming no variability is present. 
Using the cumulative distribution function (CDF) we therefore calculate the probability $P$ that the $\chi^{2}_{lc}$ value could be produced by chance. A variable source was defined as having $P< 0.001$. This sets a false detection rate at 0.1 per 124 sources searched (total source counts discussed below). 

Secondly we calculated the de-biased modulation index for all source light-curves (see \citealt{Akritas1996}; \citealt{Barvainis}; \citealt{Sadler_2006}). The modulation index (see equation 5 in \citealt{Bell_2014} ) does not take into account the errors on the flux measurements. A bias therefore exists whereby sources with large errors (or low signal to noise) have larger modulation indices. The de-biased modulation index corrects for this effect. The de-biased modulation index is defined as 

\begin{equation}
 m_{d} = \frac{1}{\overline{S}} \sqrt{\frac{\sum_{i=1}^{n}(S_{i} - \overline{S})^{2}   - \sum_{i=1}^{n} \sigma_{i}^{2}  }{n}}, 
\label{debiased_mod_index}
\end{equation}

\noindent where $\overline{S}$ is the mean of the flux density measurements. \noindent When $m_{d}$ is imaginary the magnitude of the variability is considered uncertain, due to the size of the errors $\sigma_{i}$ (see \citealt{Barvainis}). If $m_{d}$ is imaginary we evaluate the modulus of the function inside the square root. We then multiply the final expression by minus one i.e.  

\begin{equation}
 m_{d} = -1 \times \Bigg[ \frac{1}{\overline{S}} \sqrt{ \left| \frac{\sum_{i=1}^{n}(S_{i} - \overline{S})^{2}   - \sum_{i=1}^{n} \sigma_{i}^{2}  }{n} \right| } \Bigg]. 
\label{debiased_mod_index_v2}
\end{equation}

\noindent This allows us to compute and plot all values for $m_{d}$.  

For comparison with \cite{Carilli} and \cite{Mooley_2013} we also calculate the fractional variability, defined as:

\begin{equation}
\Delta S = \frac{S_{max} - S_{min}}{\overline{S}}
\label{frac_change}
\end{equation}

\noindent where $S_{max}$ and $S_{min}$ are the maximum and minimum flux density measurements for a given source. We classify a highly variable source as having $\Delta S > 50$\%.

In this dataset we classify a variable as a source that was detected in all epochs and showed significant variability (based on the $\chi^{2}$ statistic). We classify a transient source as one that was detected in less than three epochs, with large signal to noise ratio. 
The distinction is fairly trivial but the physical processes driving the different classes may possibly be different e.g. explosive and one off vs. intrinsic and persistent. In conjunction with the variability statistics described above we inspected all sources that were not detected in all three epochs. 

\section{Source counts}
In total 124 Gaussian source components were detected in all three images. Table~\ref{obs_table} gives the number of blind detections as a function of epoch. Depending on the image and noise level different Gaussian components were blindly detected in extended objects. A total of 115 sources were blindly detected in the deepest image (epoch~1). A further nine Gaussian components were detected in epochs 2 and 3 (and not epoch~1) due to the complex morphology of extended objects. 

In the analysis of \cite{CDFS_paper} 142 source components are reported in epoch~1. The discrepancy between the source counts in epoch~1 (115 vs. 142) can be attributed to different source finding algorithms and source extraction levels. The VAST pipeline represents each source as a combination of Gaussian components. Unresolved or partially resolved sources are characterised as  single elliptical Gaussians, whilst resolved sources are characterised by a number of such components. Over all three epochs, for each blindly detected Gaussian fit, a complete light-curve was produced (using forced measurements if the source was undetected in one or more epochs). These light-curves were then tested for significant variability (as described in section~4). 

\section{Results}
\subsection{Variable sources}
\begin{figure}
\centering
\includegraphics[scale=0.52]{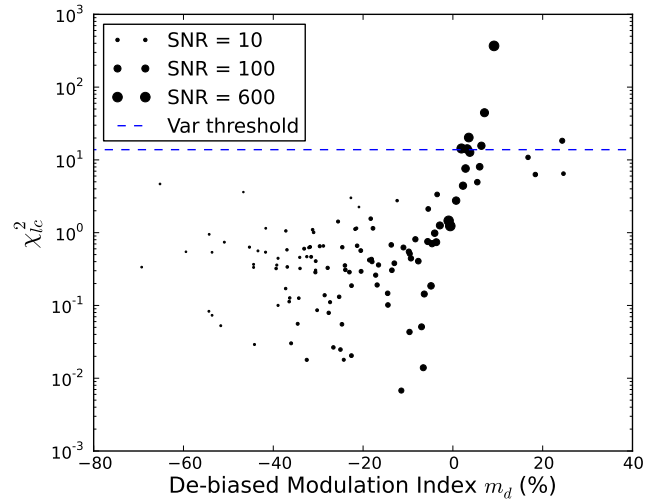}
\caption{A plot of the de-biased modulation index versus $\chi_{lc}^{2}$ for all sources in the sample. The size of the marker is proportional to the mean signal-to-noise ratio of the source detection. The dashed line denotes the variability detection threshold. Seven sources sit above the variability detection threshold. Four of these sources are considered genuine variables and are used in the subsequent analysis.}
\label{chi_vs_v2}
\end{figure}

In Figure \ref{chi_vs_v2} we plot the variability statistics as a function of signal-to-noise (SNR) ratio for all of the sources in the sample. 
The large amount of sources with negative de-biased modulation indices is a result of having two epochs that are less sensitive. The source finder is very efficient at detecting objects in epoch 1, which are subsequently measured in the less sensitive epochs (2, 3) with slightly larger error bars. To check that this was expected we wrote a simulation. Using a Gaussian number generator we simulated the light curves of 124 sources which had error bars in proportion to those from the three epochs. For this simulation the de-biased modulation index shifted to be predominantly negative for sources with low SNR. We then repeated the calculation, this time keeping the error bars (for the three data points) of equal size. This yielded an even distribution of modulation indices around zero.    

\begin{figure*}
\includegraphics[scale=0.55]{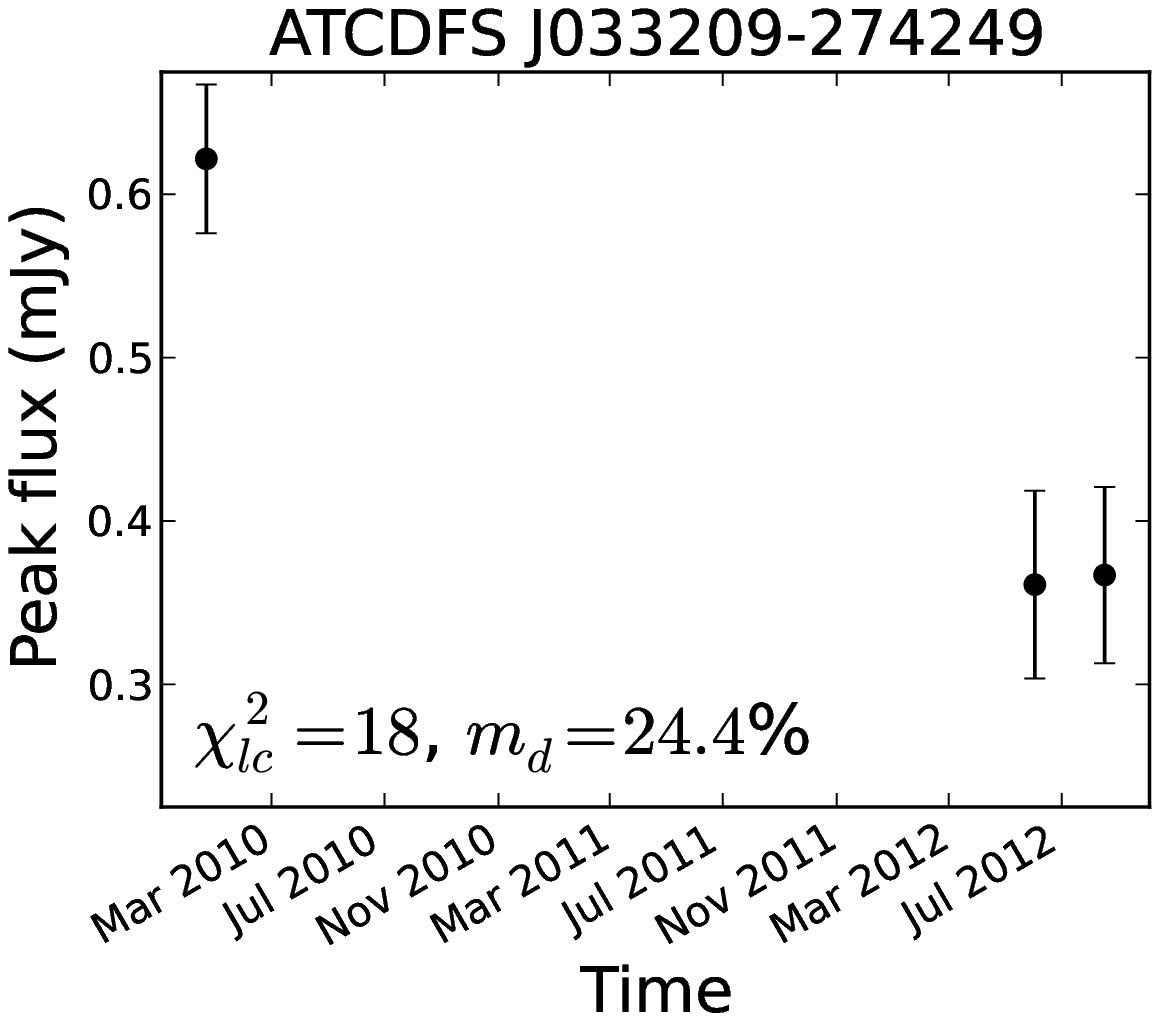}
\includegraphics[scale=0.55]{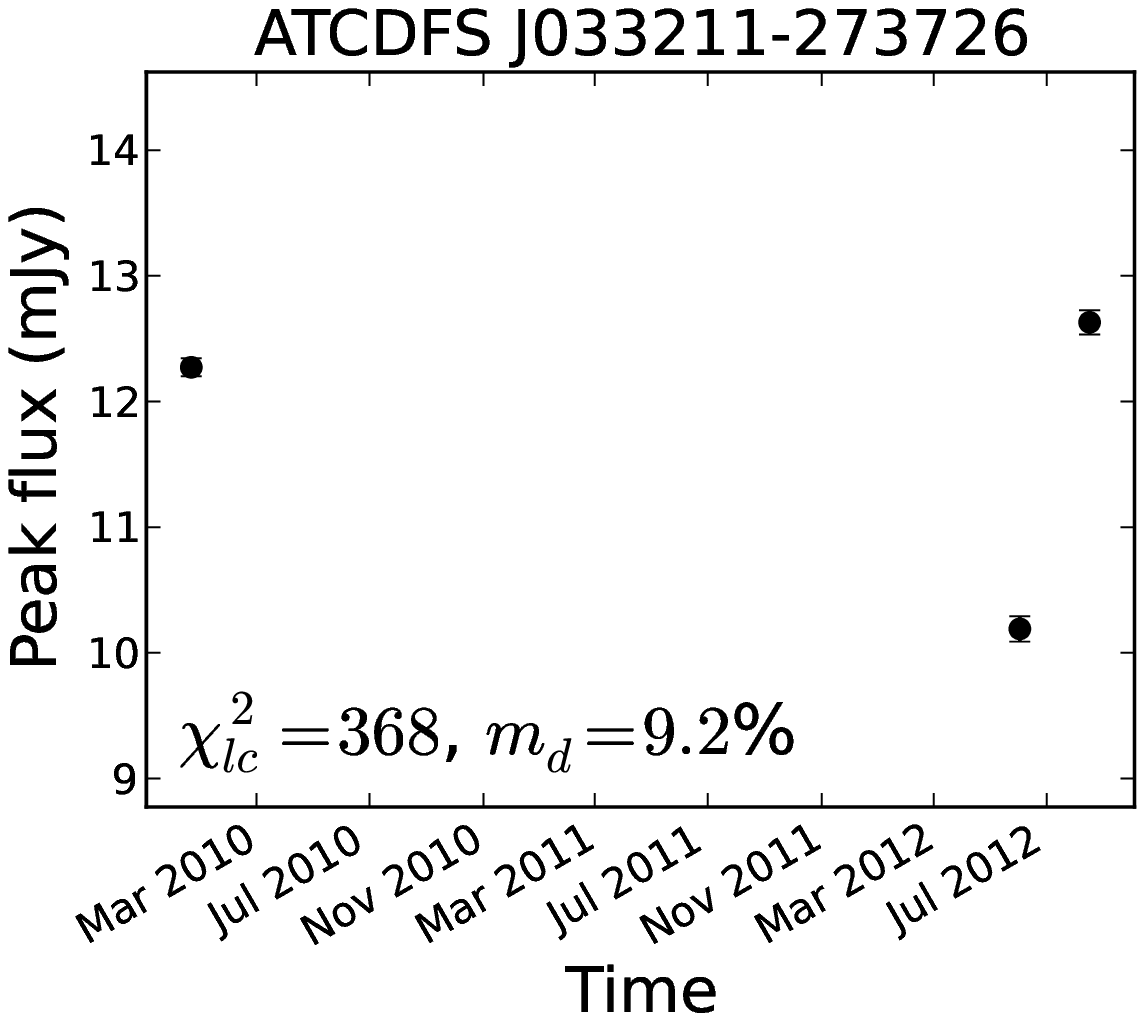}
\includegraphics[scale=0.55]{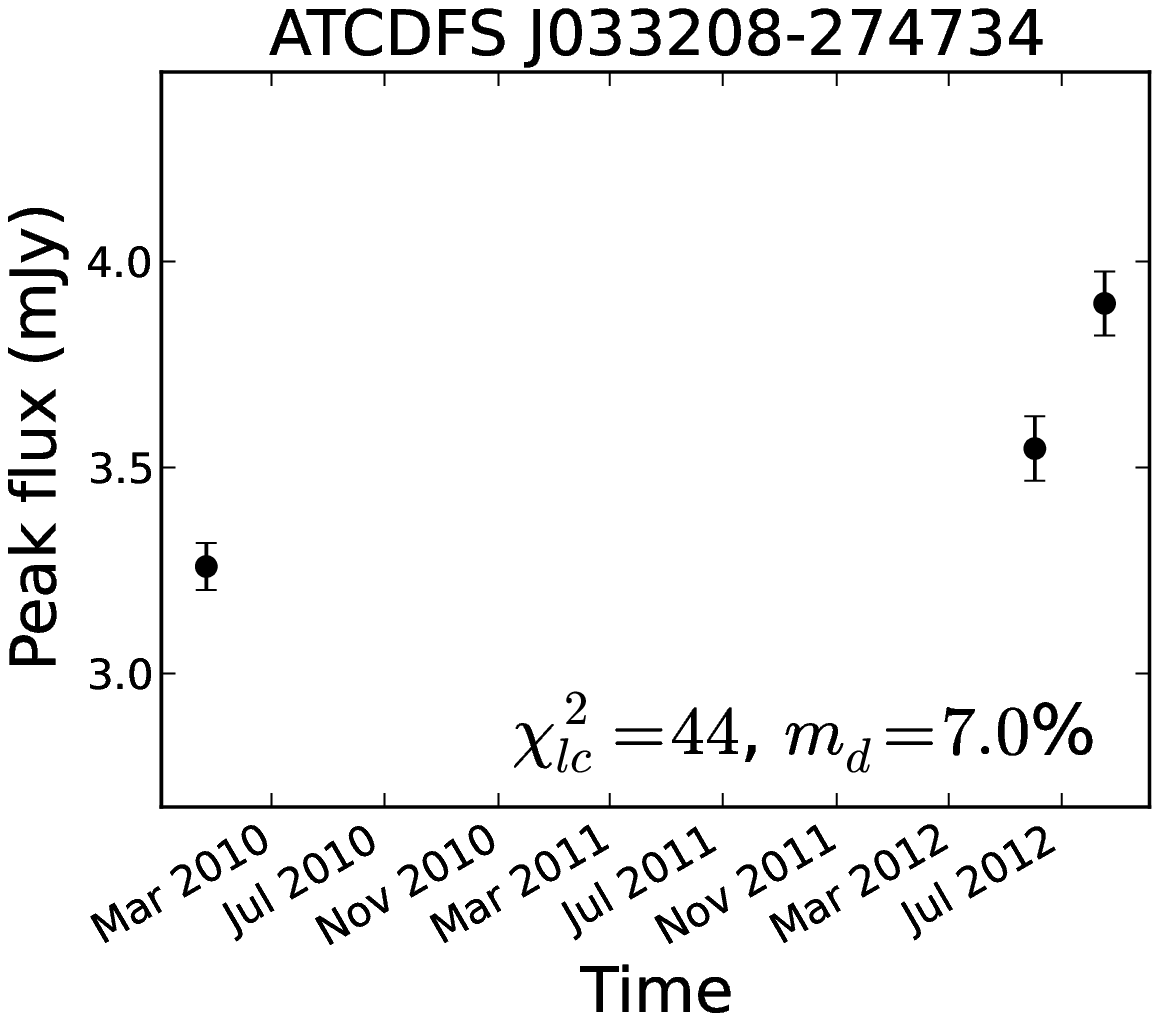}
\includegraphics[scale=0.55]{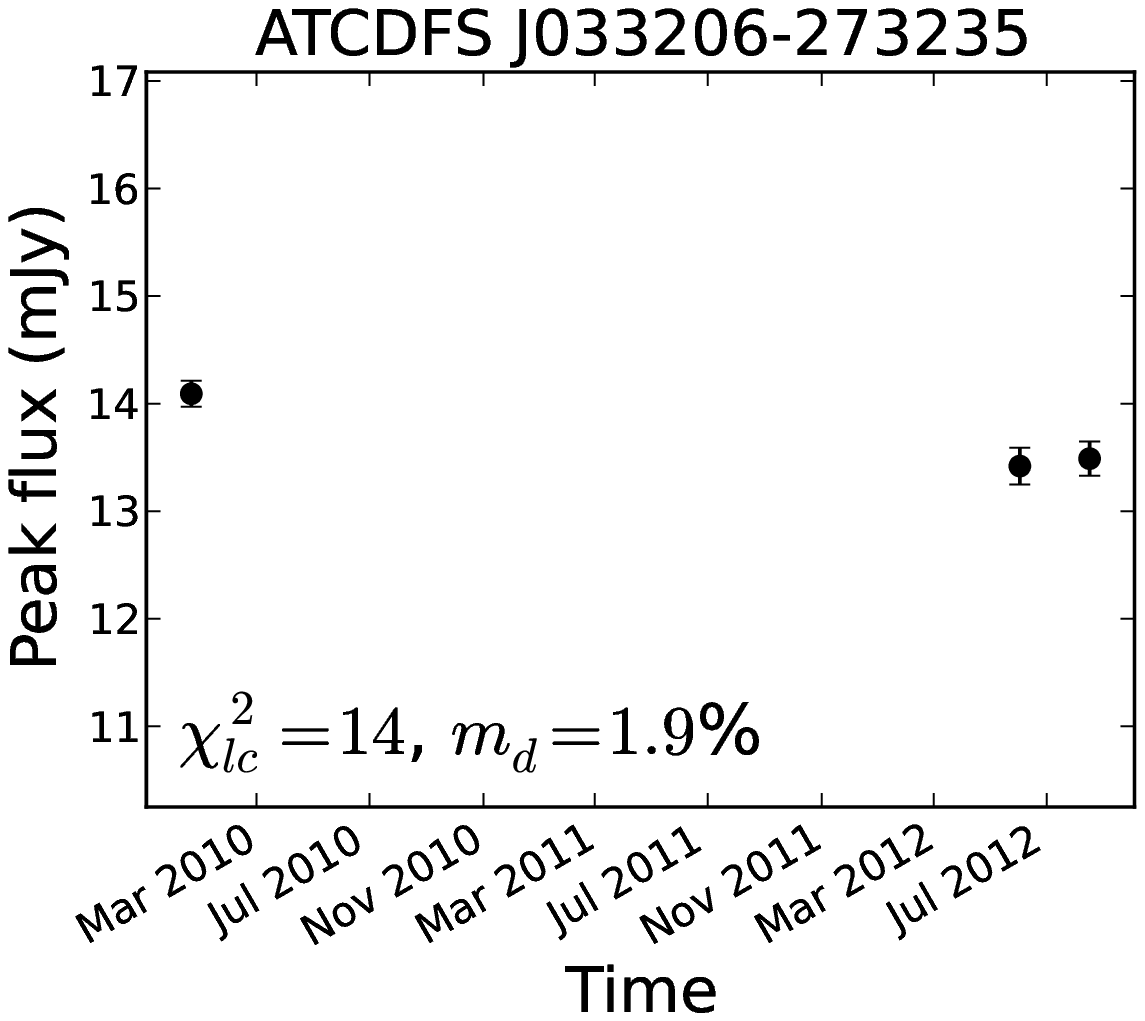}
\caption{Light-curves of the variable sources listed in Table \ref{var_table}. For each of the sources, the $\chi^{2}_{lc}$ and de-biased modulation index $m_{d}$ are indicated in the plot. The top left light-curve has a y-scale range of $\pm50\%$ of the mean flux. The remaining light-curves have a y-scale range of $\pm25\%$.}
\label{LCs}
\end{figure*}

In total seven sources were identified which met our criteria of being variable. After visual inspection one of the sources was found to be part of an extended multi-component object and we discarded the variability because the Gaussian fits were confused, extended and unreliable between images. Two further sources were discarded because the morphologies were extended and the de-biased modulation indices were low. Four unresolved sources were considered to have significant astrophysical variability and we will discuss the individual properties of these below, also see Figure \ref{LCs}. The properties of these variable sources are summarised in Table~\ref{var_table}.   

\begin{table*}
\centering
\caption{Properties of the variable sources. The $\overline{SNR}$ column denotes the mean signal-to-noise ratio of the flux measurements. The $\alpha^{5.5GHz}_{1.4GHz}$ column gives the spectral index (where $S \propto \nu^{+\alpha}$) measured between the 1.4 GHz flux density (taken from \citealt{Miller}) and 5.5 GHz averaged data from all three epochs (Huynh et al. in prep). See Huynh et al. (2012) for positional errors on individual source positions.}
\begin{tabular}{|c|c|c|c|c|c|c|c|c|c|c|}
\hline
Name &RA (J2000)  & Dec (J2000) & $\overline{SNR}$ &$\chi^{2}$ & $p$-value & $m_{d}$ (\%)& $\Delta S$ (\%) & $\alpha^{5.5GHz}_{1.4GHz}$ & Host  \\
\hline
ATCDFS J033209-274249  & 3h32m9.7s & -27d42m48.4s & 29.3 & 18.3 & $1.0\times 10^{-4}$ & 24.4 & 57.9 & +0.53 & AGN \\
ATCDFS J033211-273726 & 3h32m11.7s & -27d37m26.3s & 629.2 & 368.5 &  $<1\times 10^{-15}$ & 9.2 & 20.9 & +0.83 & QSO+AGN \\
ATCDFS J033208-274734 & 3h32m8.6s & -27d47m34.6s & 217.3  & 44.5 & $2.2\times 10^{-10}$ & 7.0 & 17.9 & +0.41 & QSO+AGN\\
ATCDFS J033206-273235 & 3h32m6.1s & -27d32m35.8s & 428.4 & 14.3 & $7.8\times 10^{-4}$ & 1.9 & 4.9 & +0.15 & AGN \\
\hline
\label{var_table}
\end{tabular}
\end{table*}

\subsubsection{ATCDFS J033209-274249}

This source showed the largest de-biased modulation index $m_{d}=24.4$\% (see Figure \ref{LCs} upper left panel) and fractional variability $\Delta S = 57.9$\%  and it was the faintest of the four variables. This source was not identified as being variable in the \cite{Mooley_2013} sample. It has a spectroscopic redshift of $z = 0.733$ (Szokoly et al. 2004) and is classified as a type-one AGN from its X-ray properties (Afonso et al. 2006). The HST ACS F180LP image from CANDELS (Grogin et al. 2011, Koekemoer et al. 2011, Skelton et al. 2014) shows a spheroidal optical host with a nearby companion to the south west  (see Figure \ref{optical}, top left).  This companion has a photometric redshift of 0.73 (Ferreras et al. 2005). The variable source and its companion appear to be members of a cluster at $z = 0.734$ (Gilli et al 2003, Szokoly et al. 2004, Silverman et al. 2010). 

\begin{figure}
\includegraphics[scale=0.38]{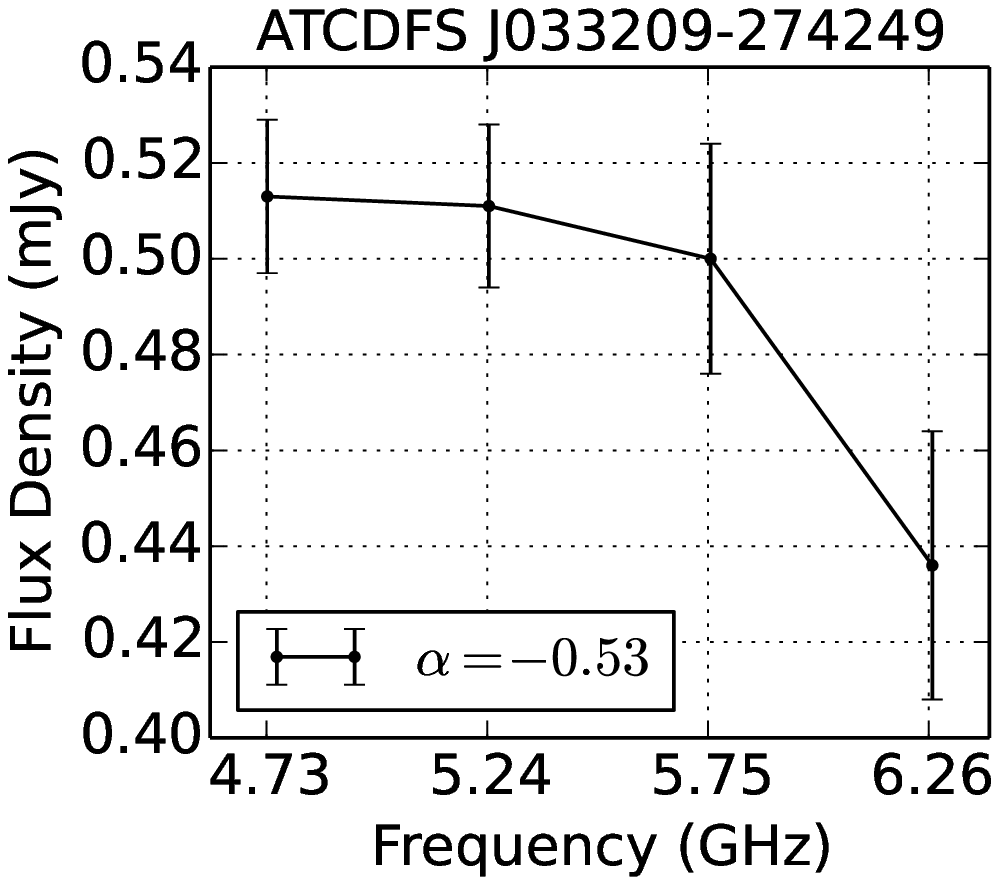}
\includegraphics[scale=0.38]{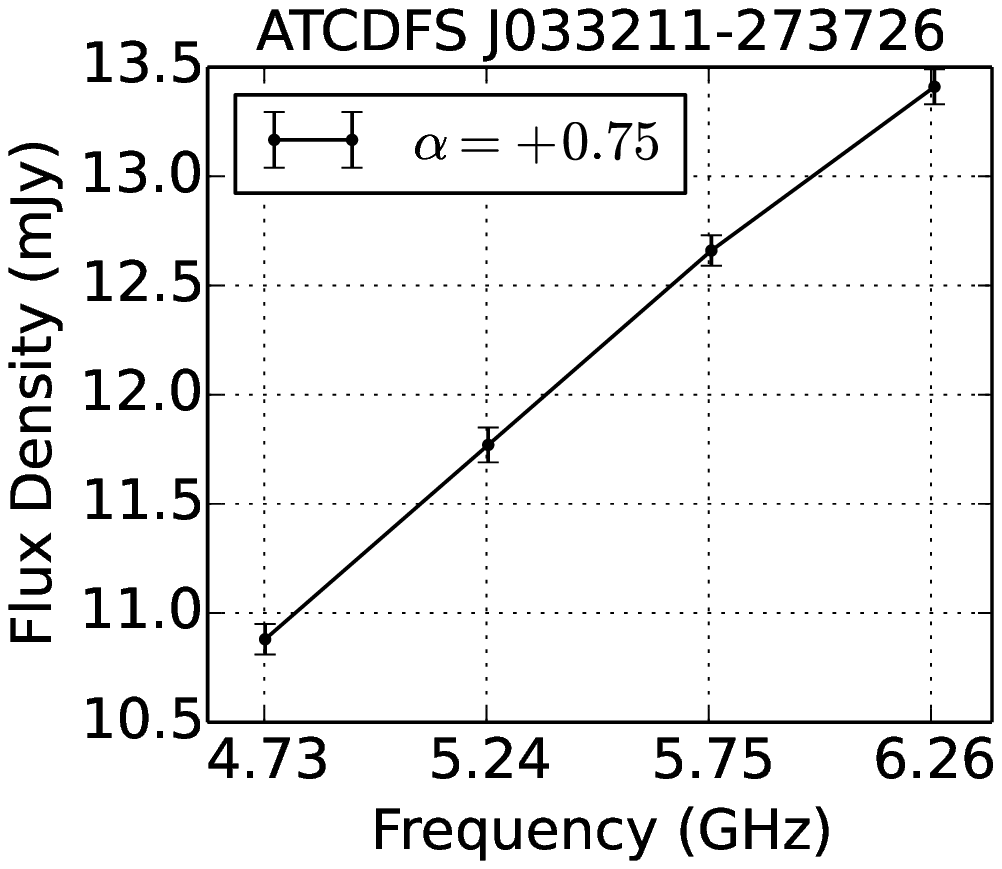}\\
\includegraphics[scale=0.38]{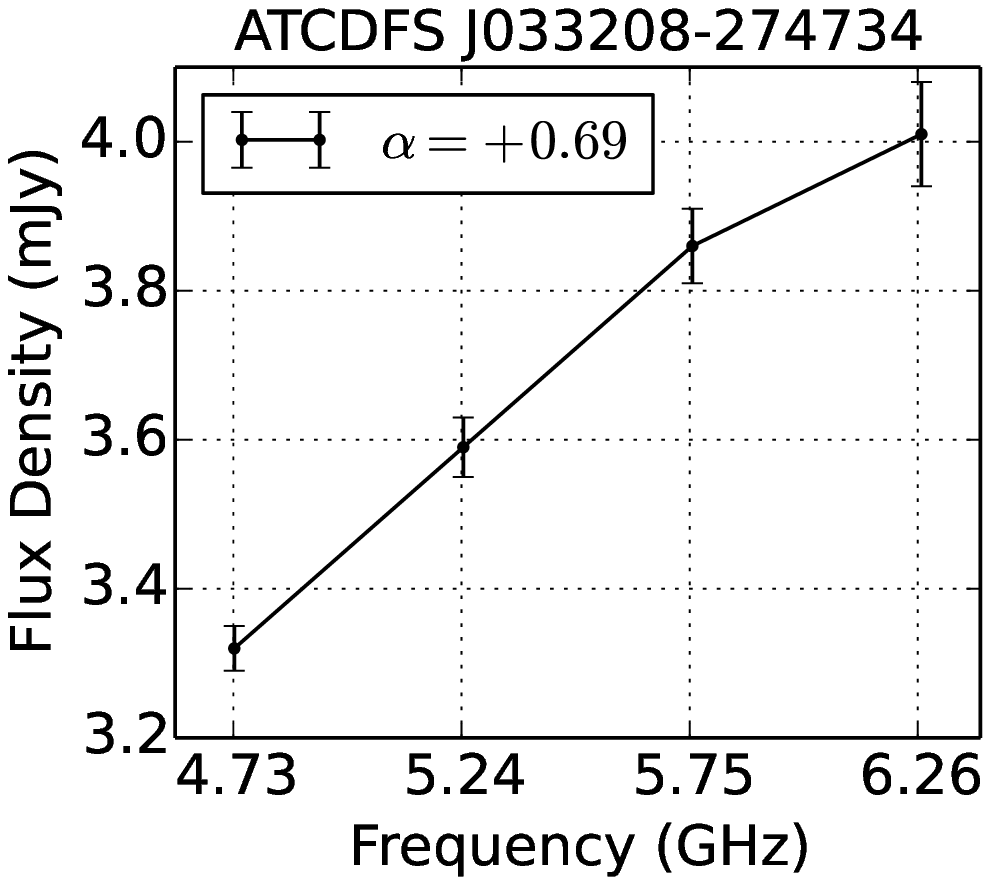}
\includegraphics[scale=0.38]{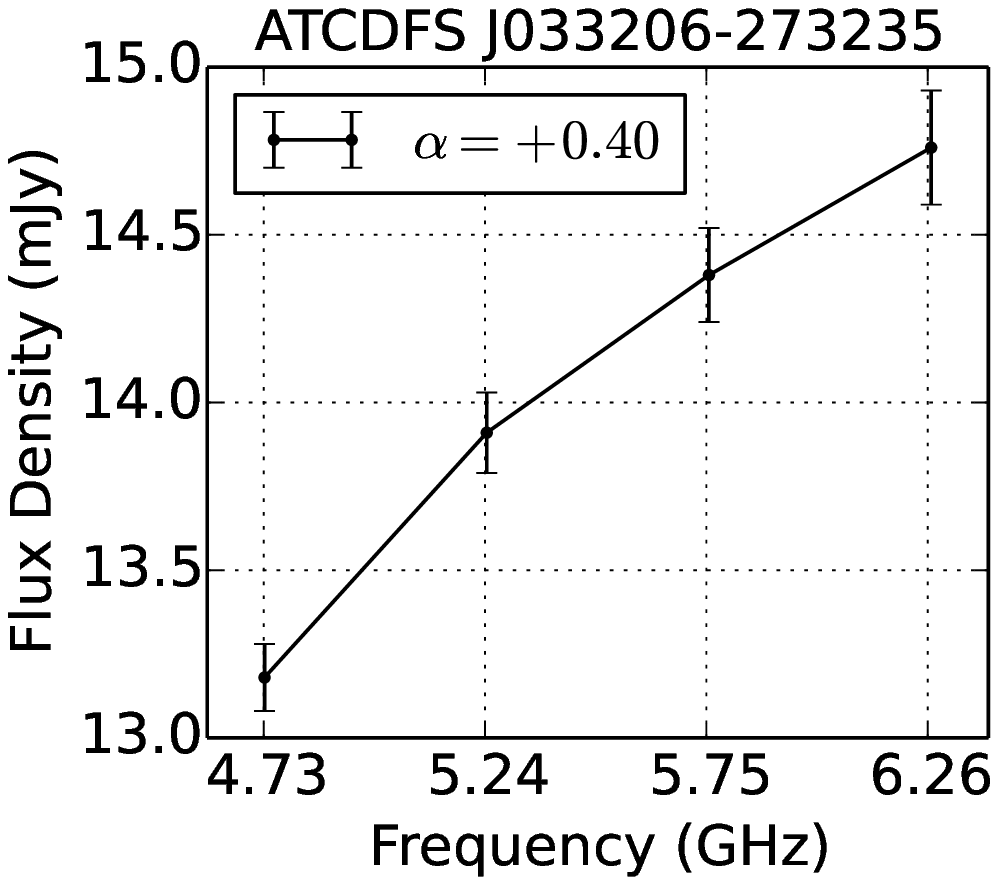}
\caption{Intra-band spectral energy distributions for the four variable sources. These spectra are measured in a deep image of all epochs (Huynh et al. in prep) across the 2~GHz observing bandwidth which was centred at 5.5~GHz.}
\label{spectral_index}
\end{figure}

This source has an inverted radio spectrum with $\alpha^{5.5GHz}_{1.4GHz}$ = +0.52 where $S \propto \nu^{+\alpha}$ (Huynh et al. in prep). This spectral index was calculated between archival measurements by \cite{Miller} at 1.4~GHz and measurements in a deep image of all three epochs at 5.5~GHz (Huynh et al. in prep). The spectral indices between 1.4~GHz and 5.5~GHz for the individual epochs are $\alpha_{1}=+0.65$, $\alpha_{2}=+0.29$ and $\alpha=+0.29$. 
The intra-band spectral index is relatively flat until 5.8 GHz, above which point it becomes negative (see Figure~\ref{spectral_index}). The intra-band spectral index is measured in a deep image of all three epochs (Huynh et al. in prep). 

\subsubsection{ATCDFS J033211-273726}

This source had the highest $\chi^{2}_{lc}$ value (lowest $P$-value) of all the variables and a de-biased modulation index $m_{d}=9.2$\% (see Figure \ref{LCs} upper right panel). A significant dip in flux density is observed in epoch-2 with respect to epochs 1 and 3.  The source had previously been identified by \cite{CDFS_paper} to be variable with respect to archival measurements by \cite{VLA_CDFS} and we confirm that continued radio variability via this survey. This source was also identified as being variable at 1.4 GHz by \cite{Mooley_2013}. This source has an R-band (Vega) magnitude of 18.626 and is classified as having a QSO-like optical SED \citep{wolf_2008}. An optical spectrum of the source shows broad emission lines indicative of quasar activity and places the source at redshift $z=1.575$ \citep{silverman_2010}. 

The radio spectral index is $\alpha^{5.5GHz}_{1.4GHz}$ = +0.83 (Huynh et al. in prep), so it is an inverted spectrum source. The intra-band spectral index is $\alpha^{6.2GHz}_{4.7GHz}$ = +0.75, see Figure~\ref{spectral_index}. The broad optical emission lines and inverted radio spectrum suggests a line of sight relatively face-on to the torus of the AGN with a view to the accretion disk. This is an X-ray luminous source with L(0.2 -- 2 keV) = $2 \times 10^{44}$ erg s$^{-1}$, classifying it as an AGN \citep{xue_2011}. Figure \ref{optical} (top right panel) shows the compact optical counterpart in the HST ACS F180LP image \citep{rix_2004}, confirming the QSO nature. 

\subsubsection{ATCDFS J033208-274734}
The light-curve of this source (Figure \ref{LCs} bottom left panel) shows an increase in flux over all three epochs with $m_d=7.8$\%. This source was not identified as being variable in the \cite{Mooley_2013} sample. It has been classified as a QSO using the ultraviolet excess method (Croom et al. 2001) and it also has a QSO-like optical SED (Wolf et al. 2008). The optical spectrum shows emission lines indicative of an AGN at z = 0.544 (Croom et al. 2001). This source is classified as a type-one QSO from its X-ray luminosity and hardness ratio (Szokoly et al. 2004). 

\begin{figure*}
\includegraphics[scale=0.75]{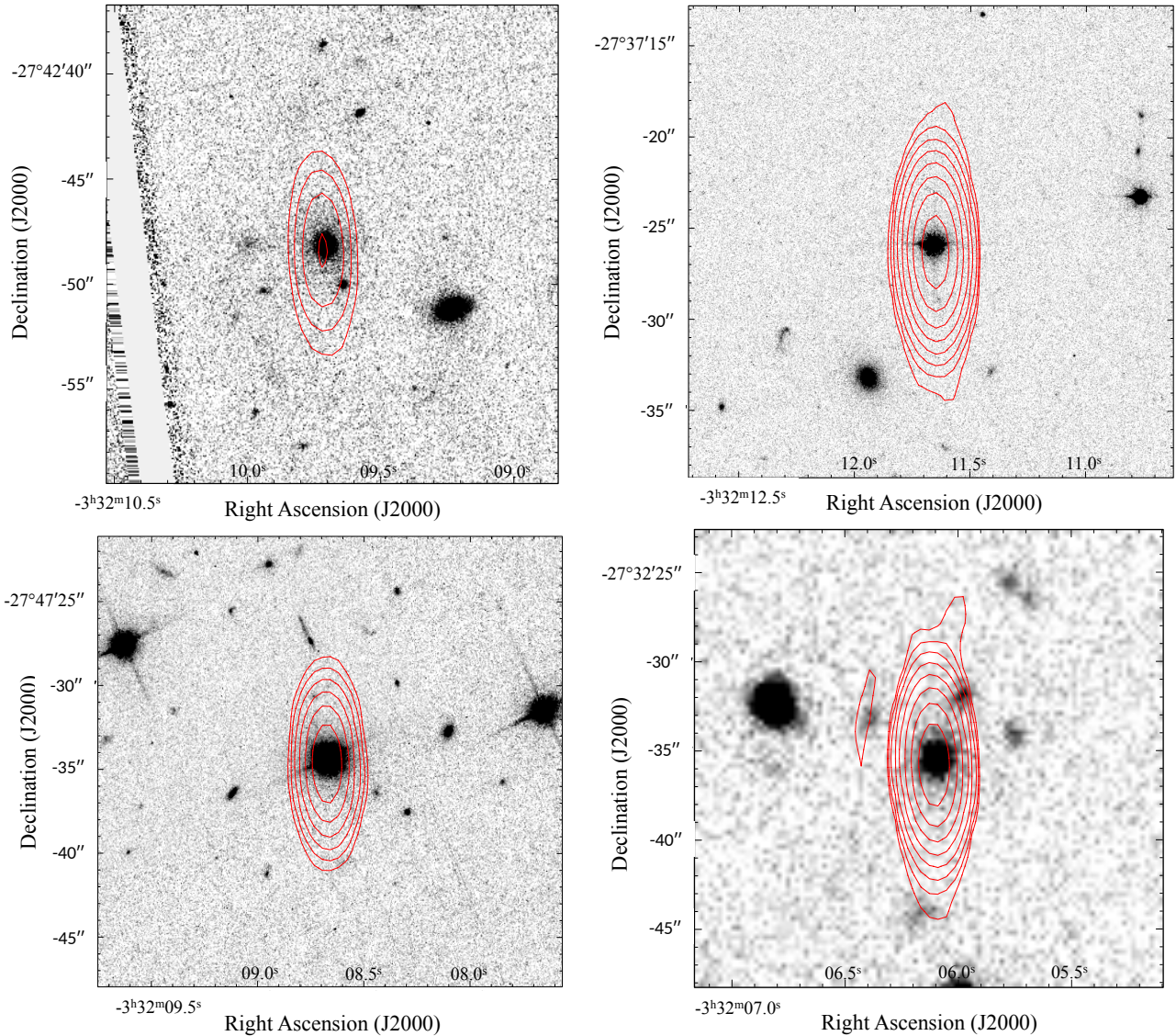}
\caption{Optical HST images of the variable sources ATCDFS J033209-274249 (top left), ATCDFS J033211-273726 (top right) and ATCDFS J033208-274734 (bottom left). These panels are HST ACS F180LP images from CANDELS (Grogin et al. 2011, Koekemoer et al. 2011, Skelton et al. 2014). The bottom right panel shows a MUSYC i band image (\citealt{MUSYC}) of ATCDFS J033206-273235. The radio contours from this work start at five times the image noise and increase by multiples of two.}
\label{optical}
\end{figure*}

The source has an inverted radio spectrum $\alpha^{5.5GHz}_{1.4GHz}$ = +0.41 (Huynh et al. in prep). It has an intra-band spectral index of $\alpha^{6.2GHz}_{4.7GHz}$ = +0.69, see Figure \ref{spectral_index}. This source is point-like in HST ACS F180LP image from CANDELS (Grogin et al. 2011, Koekemoer et al. 2011, Skelton et al. 2014) (see Figure \ref{optical}, bottom left). This source was also identified as an optical variable by \cite{trevese} and potentially the radio and optical variability are intrinsically linked e.g. see \cite{elme}. 

\subsubsection{ATCDFS J033206-273235}

This source showed a very low level of variability $m_{d}=1.9$\% (see Figure \ref{LCs} bottom right panel) and was the brightest of the variable sources and all other sources in the field. It was not identified as being variable in the \cite{Mooley_2013} sample, but the NVSS 1.4 GHz flux density differs from the later epoch eCDFS observations by 3 mJy (almost 40\%). Regardless of the low de-biased modulation index the variability is considered significant and also the radio spectral index is intriguing, so we include it in our variables sample.  

It has a reliable spectroscopic redshift of $z = 0.961$ (Silverman el. al. 2010), with the spectrum only showing narrow emission lines. It is formally unresolved in our 5.5 GHz radio imaging but marginally resolved in higher resolution VLA 1.4 GHz imaging (Miller et al. 2013). Its X-ray luminosity L(0.2 -- 2 keV) of $3 \times 10^{43}$ erg s$^{-1}$ (Vattakunnel et al. 2012) places it in the regime of X-ray luminous AGN (e.g. Bauer et al. 2004). This source has an inverted radio spectrum with $\alpha^{5.5GHz}_{1.4GHz}$ = +0.15 and intra-band spectral index of  $\alpha^{6.2GHz}_{4.7GHz}$ = +0.40, see Figure \ref{spectral_index}.

\subsection{Transient sources}
No transient radio sources were found above 5.5$\sigma$. As discussed in section 4, by using our forced measurement algorithm and inspecting light-curves with less than three blind detections we would have been capable of detecting any such transient source above this limit within this dataset. 
\section{Discussion}
\subsection{Transient sources}
No radio transients were detected in this survey. We can therefore place upper limits on the surface density of events. Using Poisson statistics (see \citealt{gehrels}) with zero detections at the 95\% confidence level we can set an upper limits on the surface density ($\rho$) of events per deg$^{2}$ as: 

\begin{equation}
\rho < \frac{\ln(0.05)}{(n-1) \times \Omega}, 
\label{upp_eq}
\end{equation}

\noindent where $n$ is the number of epochs and $\Omega$ is the sky area surveyed (deg$^{2}$). We use the sky area and corresponding mean sensitivities defined in Table \ref{area_sens_table} to calculate the upper limits on the surface density given in Table \ref{upper_lims}. Within a region of 0.2 deg$^{2}$, where the sensitivity profiles of the images are fairly flat (see Figure \ref{area_vs_sens}), we find an upper limit $\rho < 7.5$~deg$^{-2}$ above 68.8~$\mu$Jy on timescales of 2.5 months and 2.5 years at 5.5~GHz. \\

\begin{table}
\centering
\caption{Upper limits on the surface density of transient sources on timescales of 2.5 years and 2.5 months, as a function of detection threshold. The detection thresholds are defined as the mean sensitivities given in Table \ref{area_sens_table} multiplied by the source extraction level of 5.5($\sigma$).}
\begin{tabular}{|c|c|c|}
\hline
Cumulative Area  & Detection threshold & Surface density $\rho$   \\
 (deg$^{2}$) &  ($\mu$Jy beam$^{-1}$) & (deg$^{-2}$) \\
\hline
0.1 & 64.4 & $<$15.0  \\
0.2 & 68.8 &  $<$7.5 \\
0.3 & 82.3 &   $<$5.0 \\
0.4 & 143.0 &   $<$3.8 \\
0.5 & 346.5 &   $<$3.0 \\
\hline
\label{upper_lims}
\end{tabular}
\end{table}

These calculated surface densities are dependent on the transient evolution timescale. The cadence of these observations are not explicitly folded into equation~\ref{upp_eq} but they are implied by $n$ i.e the number of epochs that are separated by a given cadence, or in this case, two different timescales (2.5 years and 2.5 months). For incoherent sources the typical evolution timescale is $ \lesssim 100$~days (\citealt{metzger}; \citealt{Pietka}), which is well matched by our 2.5 month timescale (or 77 days). The quoted surface densities of transients therefore may be even lower for long duration incoherent sources (e.g. 2.5 years), as the long duration sampling is not well matched to the evolution timescale (for this example).   

\subsection{Comparison with previous transient surveys}
In comparison to previous work, \cite{Bower_2007} report no radio transients on the timescale of one year, over 17 individual epochs, with a typical flux density threshold of 90~$\mu$Jy (at the half-power point). This equates to an upper limit of $< $2.1 deg$^{-2}$ at the 95\% confidence level (using equation~\ref{upp_eq} assuming a field of view of 0.09~deg$^{2}$ and 17 epochs). This upper limit is shown in Figure \ref{surface_density} and is more constraining than our upper limits at comparable detection levels. 


\cite{Bell_2011} searched for transients in 24 years of archival Very Large Array at 1.4, 4.8 and 8.4~GHz. An upper limit of $\rho<0.032$~deg$^{-2}$ was placed on the surface density of transients $>8$~mJy with typical timescales of 4.3 to 45.3 days. \cite{Bell_2011} were sensitive to shorter timescale transient events at a higher detection threshold over a range of frequencies. It is therefore difficult to compare the survey presented in this paper with the results of \cite{Bell_2011}. 

\cite{Ofek_2011} report the possible detection of a transient at 5 GHz with a peak flux density of 2.4~mJy. The transient was detected in only one epoch at the start of the survey, which sampled timescales of days to years. Because the transient was detected in the first epoch of the survey the duration was uncertain. \cite{Ofek_2011} report that a sky surface density of transients 0.039~deg$^{-2}$ is expected above 1.8~mJy. 

Scaling this upper limit to a detection threshold of 0.1~mJy (assuming a euclidean population with $\rho \propto S_{v}^{-1.5}$, where $S_{v}$ is the detection threshold) we would expect to find $\rho<3.0$~deg$^{-2}$, or 0.89 sources over our field of view. This is just below the upper limits on the surface density of transients placed through this work. It is therefore unlikely that we would have been able to detect a transient of this type.

\subsection{Variable sources}
We find four variable radio sources on timescales of months to years. With a total of 124 sources this equates to 3\% of the sample that were variable. All of the variable sources are contained within a region covering 0.3~deg$^{2}$. Due to the high signal to noise ratio of the source detections and relatively flat noise profile of the images, all four variables would have been detectable within any part of the 0.3~deg$^{2}$ region. This gives a surface density of 13.3 variable sources per deg$^{2}$.   

All four of the variable sources had positive spectral indices between the average 5.5~GHz measurements and 1.4 GHz measurements taken in 2007 by Miller et al. (2013). ATCDFS J033211-273726 had a spectral index of $\alpha = +0.83$. For this source over year long timescales we measure a fraction variability of 20.9\% (see table \ref{var_table}). Adjusting the average 5.5~GHz flux density accordingly we can place errors on the spectral index of $\alpha = 0.83^{+0.17}_{-0.13} $ on this timescale. Therefore if the current level of 5.5~GHz variability is typical, it is difficult to reconcile this source as having been flat spectrum, or even negative in the recent past (assuming the 1.4 GHz flux has remained relatively constant). The intra-band spectral indices confirm the inverted spectra for three of the variable sources (see Figure \ref{spectral_index}).

The inverted radio spectra combined with the optical and X-ray identifications suggest that these are compact sources, that are possibly young and have active radio jets. Synchrotron self absorption can cause inverted radio spectra during outbursts, or shocks, in relativistic jets (see \citealt{MG1985}). Typically the outburst is followed by an optically thin phase resulting in a negative spectral index (at the reference frequency of 5.5~GHz). The optically thin phase is a function of frequency and it will evolve from high to low frequencies as a function of time. This model seems most applicable to ATCDFS J033209-274249 i.e. the intra-band spectral index is fairly flat, and we could interpret the light-curve and inverted archival spectral index to suggest a recent outburst. The spectrum of this source also evolves from $+0.65$ to $+0.29$. This supports the shock-in-jet model because the higher frequency end of the intra-band spectrum is negative and the averaged spectral index per epoch is evolving to become negative .  

Alternatively these are young radio sources with gigahertz peaked-spectrum (GPS). The variability could therefore be related to relatively recent jet activity. Variability has been reported before in GPS sources by, for example, \cite{Bolton_2006}. In this study they note that the majority of sources that had their spectra peak below 5~GHz were non-variable, whereas almost all the sources which had their peak above 5~GHz were variable. The variable sources ATCDFS J033211-273726, ATCDFS J033208-274734 and ATCDFS J033206-273235 have peaks in their spectra above 5~GHz and may well be explained by GPS variability.   

The abundance of GPS sources is difficult to quantify for a sample at a given flux level and frequency. Typically the turnover in a gigahertz peaked source spectrum can vary from 500~MHz up to 10~GHz or more \citep{odea}. \cite{odea} report that the expected abundance of GPS sources within a sample greater than 1~Jy is $\sim$10\% at 5~GHz. \cite{CDFS_paper} report that of the 108 eCDFS sources (non multi-component) that had reliable 1.4~GHz crossmatches, 10 (or 9\% of the sample) had spectra $\alpha^{5.5}_{1.4}>+0.5$. This compares with 3\% of the sources that we find to be variable and have inverted spectral indices. Clearly not all of the steep spectrum sources are variable within our sample, however, this assumes that we were capable of detecting the variability in the weaker GPS sources (see discussion in section 7.5). Given the expected abundance of GPS sources in this sample, these variables do constitute a fairly rare region of the spectral population. 

We crosschecked the seven variables reported in the full variability analysis of \cite{Mooley_2013} with our sample. ATCDFS J033211-273726 (discussed above) was the only source found to be variable in both surveys. The difference between this survey and \cite{Mooley_2013} was the typical timescale of the observations and the frequency (1.4~GHz vs. 5~GHz). \cite{Mooley_2013} sampled typical timescales of one day to three months whereas this survey samples $\sim$2.5 months and $\sim$2.5 years. The timescales for relativistic jets to evolve at these frequencies will be longer than the scintillation timescale and hence will be better selected by our observing strategy (i.e. timescale and frequency) when compared to \cite{Mooley_2013}. 

\cite{Mooley_2013} used the Miller et al. (2013) 1.4~GHz and \cite{CDFS_paper} 5.5 GHz catalogs to calculate the spectral indices. They report that four variables had negative spectral index, and three were positive (with a range $-0.4$ to $+0.89$). Flat spectrum sources are often compact and can undergo refractive scintillation, therefore the difference in spectral properties between the variables in our sample, and those in \cite{Mooley_2013}, may be influenced by the presence of scintillation.  

\subsection{Comparison with previous variability surveys}
\begin{figure*}
\centering
\includegraphics[scale=0.60]{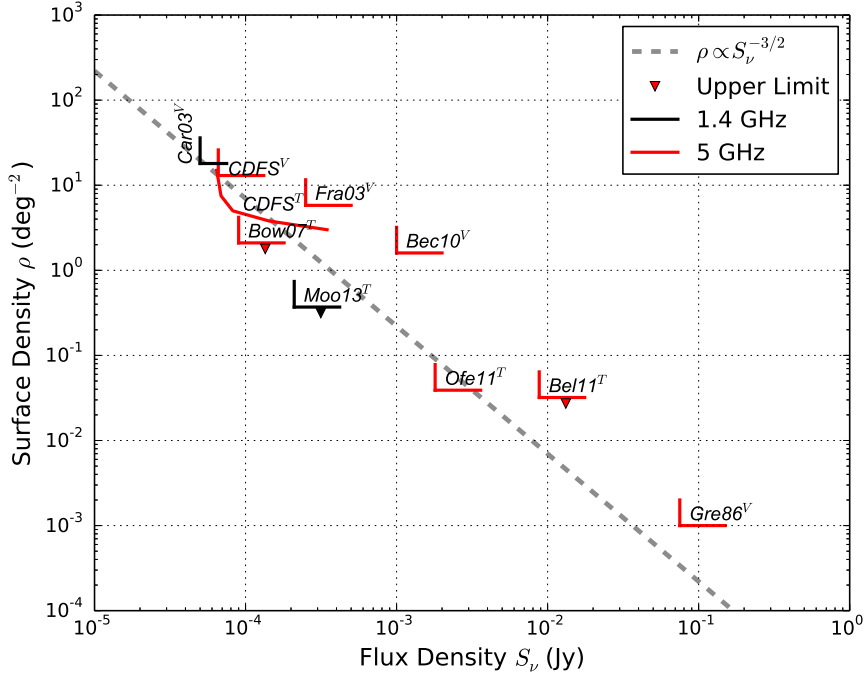}
\caption{Surface density (deg$^{-2}$) of transient and variable sources versus flux density (Jy) for a selection of surveys. This plot is focused primarily on surveys at 5~GHz but also includes relevant surveys at other frequencies. Superscript V and T indicate variable and transient surface densities respectively. \textit{CDFS$^{V}$} gives the surface density of 13~deg$^{-2}$ at a flux density $>82.3$~$\mu$Jy for all the variables detected in this paper. \textit{CDFS$^{T}$} gives the upper limits on surface density of transients calculated through this work. The dashed grey line shows a fit normalised to the CDFS$^{V}$ surface density where $\rho \propto S_{\nu}^{-3/2}$. This assumes that the variable source population is drawn from a homogenous euclidean population. 
The additional surveys are labelled accordingly. 
\textit{Gre86$^{V}$} - Gregory \& Taylor (1986);
\textit{Car03$^{V}$} - Carilli et al. (2003); 
\textit{Fra03$^{V}$} - Frail et al. (2003);
\textit{Bow07$^{T}$} - Bower et al. (2007);
\textit{Bec10$^{V}$} - Becker et al. (2010);
\textit{Ofe11$^{T}$} - Ofek et al. (2011);
\textit{Bel11$^{T}$} - Bell et al. (2011);
\textit{Moo13$^{T}$} - Mooley et al. (2013).
}
\label{surface_density}
\end{figure*}

In Figure \ref{surface_density} we compare the surface density of variables found in this survey with those reported in the literature. \cite{Carilli} report a surface density of variable sources above 100$\mu$Jy to be  $<$18 deg$^{-2}$ at 1.4 GHz on 17 month timescales (out to a radius of 7.8$^{\prime}$). We find a comparable surface density $\sim13.3$~deg$^{-2}$, if all the variables sources within our sample are considered and we ignore the change in frequency. This calculation is performed by simply dividing the number of variables found (four) by the area (0.3~deg$^{2}$).

If we assume that our results are drawn from a Poisson process, using the method of \cite{gehrels} we can calculate upper and lower limits on the surface density. For our results, assuming four detections with a survey area of 0.3~deg$^{2}$ over three epochs we find $2.3 < \rho < 15.3$~deg$^{2}$. For the \cite{Carilli} results, if we assume that eight variables were detected over three epochs with an area of $\pi(32^{\prime})^{2}=0.89$~deg$^{2}$ we find $2.2 < \rho < 8.1$~deg$^{2}$.  Our results are therefore in broad agreement with \cite{Carilli}. 

We find only one source (ATCDFS J033209-274249.0) that is considered highly variable with $\Delta S / \overline{S} > 50\%$ (see Table \ref{var_table}). This gives a surface density of highly variable sources 3.3~deg$^{-2}$ (or 1\% of our sample). Using Poisson statistics this equates to $0.1 < \rho < 8.0$~deg$^{2}$. 

The predicted number of variable sources is of course a function of flux density, cadence, field of view and frequency. The \cite{Carilli} survey had some similar properties to our survey i.e. (i) rms noise measurements (in the range 15$-$18~$\mu$Jy beam$^{-1}$); (ii) cadence (timescales of both months and years); and (iii) field of view ($\sim0.5$~deg$^{2}$). 

At low Galactic latitudes, \cite{Ofek_2011} reported that $>$~30\% of unresolved sources brighter than 1.8~mJy were significantly variable on the timescales of days (at 5~GHz). \cite{Ofek_2011} conclude that the likely cause for the short timescale variability is refractive scintillation. On the longer timescales of two years \cite{Ofek_2011} report that a much lower fraction ($\sim$2\%) of the sample is variable ($>$0.5~mJy). The findings of \cite{Ofek_2011} on long timescales are consistent with the results presented in this paper. 

On the timescales of days to 15 years, \cite{Becker} examined the variability of sources towards the Galactic plane in the flux density range 1$-$100~mJy. A surface density of 1.6~deg$^{-2}$ variables was reported (8\% of the sample). \cite{Becker} argue that 80\% of the objects in their sample were Galactic objects. The \cite{Becker} sample could therefore differ greatly from the objects in our sample, otherwise this is a significant increase. 

\cite{GandT1986} find a lower surface density of variables towards the Galactic plane. Above $\sim$75 mJy on the timescales of days to years \cite{GandT1986} report $\sim$ 2\% of the sample was variable at the $>50$\% level (surface density $<10^{-3}$). This is four times less than \cite{Becker} although at a higher flux density cut off. 

\subsection{Variability completeness}
To assess our ability to detect variability within this dataset as a function of source signal-to-noise ratio and variability amount and type, we calculate a variability completeness metric. We take the light-curves of the four variables sources and re-calculate a hypothetical $\chi^{2}_{c}$ value after scaling the source flux density measurements by a factor $\gamma$ where $0<\gamma<1$. 
By reducing the flux density of the measurements (of the variable source light-curves) the variability becomes less significant when compared with the errors, as a function of $\gamma$.  The $\chi^{2}_{c}$ variable is defined as:

\begin{equation}
\chi_{c}^{2} = \sum_{i=1}^{n} \frac{(\gamma S_{i} - \tilde{S}_{c})^{2}}{\sigma_{i}^{2}}.
\label{chisquared2}
\end{equation}

\noindent The modified weighted mean flux density, $\tilde{S}_{c}$, is defined as 

\begin{equation}
\tilde{S}_{c} = \sum_{i=1}^{n} \left( \frac{\gamma S_{i}}{\sigma_{i}^{2}} \right) /  \sum_{i=1}^{n} \left( \frac{1}{ \sigma_{i}^{2}} \right),
\label{w_mean}
\end{equation}

\noindent where $\gamma$ is the scaling term.

\begin{figure}
\hspace{-0.2in}
\includegraphics[scale=0.5]{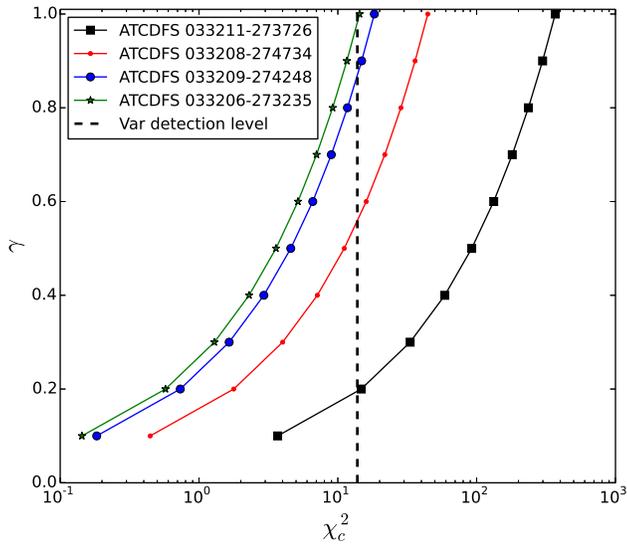}
\caption{The hypothetical $\chi^{2}_{c}$ values calculated for the four variable sources as a function of $\gamma$. The $\gamma$ factor scales the flux density measurements of the variable source light-curves (see equation \ref{chisquared2}). The dashed black line shows the variability detection threshold ($\chi^{2}>$13.9) used in this survey.}
\label{conf}
\end{figure}

Figure \ref{conf} shows the $\chi^{2}_{c}$ values as a function of $\gamma$ for the four variable sources. From Figure \ref{conf} it can be seen that ATCDFS~033211-273726, which had the highest value of $\chi^{2}_{lc}=368$, can be reduced in flux by 81\% ($\gamma=0.19$) before the variability is non-detectable. Conversely ATCDFS~033206-27273235, which had the lowest value of $\chi^{2}_{lc}=14.3$ and smallest de-biased modulation index $m_{d}=1.9\%$, could only be reduced by 2\% ($\gamma=0.98$). 

We can translate the point ($\gamma_{lim}$) at which the variability becomes non-detectable into a mean signal-to-noise ratio $\overline{SNR}_{var}$, for each of the four variable sources. We then use $\overline{SNR}_{var}$ to calculate how many other sources within the field have a $\overline{SNR}$ greater than $\overline{SNR}_{var}$. The results are summarised in Table \ref{conf_table}.

Interestingly the fairly lows levels of variability reported in this survey would only be detectable in a small number of sources. At best, the variability observed in ATCDFS~J033209-274249 could be detected in 31 other sources (or 25\% of the sample) within the field. The variability seen in ATCDFS~J033206-273235 could only be detected in four other sources (or 3\% of the sample). This is assuming that there are 124 sources within the field and some of these were multi-component or extended objects, which are unlikely to be variable. We can of course rule out 100\% variations in the majority of sources. Our statistics are influenced by the small number of epochs we have obtained (three). Acquiring more epochs should make the $\chi^{2}$ statistic more robust to detecting significant variability. 

\begin{table*}
\centering
\caption{Table summarising our ability to detect the variability found in the four variable sources within all sources in the field. $\overline{SNR}$ is the mean signal-to-noise ratio of the variable detections. $\gamma_{lim}$ is the limit on the flux scaling parameter below which the variability becomes non-detectable. $\overline{SNR}_{var}$ is the mean signal-to-noise ratio at $\gamma_{lim}$. The $N_{sources} > \overline{SNR}_{var}$ column shows how many other sources within the field have a mean signal-to-noise ratio greater than $\overline{SNR}_{var}$.}
\begin{tabular}{|c|c|c|c|c|}
\hline
Source  & $\overline{SNR}$ & $\gamma_{lim}$ & $\overline{SNR}_{var}$ & $N_{sources} > \overline{SNR}_{var}$  \\
\hline
ATCDFS J033211-273726 & 629.2 & 0.19 & 119.5 & 12  \\
ATCDFS J033208-274734 & 217.3 & 0.55 & 119.5 & 12  \\
ATCDFS J033206-273235 & 428.4 & 0.87 & 372.7 & 4  \\
ATCDFS J033209-274249 & 29.3 & 0.98 & 28.7 & 31  \\
\hline
\label{conf_table}
\end{tabular}
\end{table*} 


\section{Conclusion}
We have surveyed the extended \emph{Chandra} Deep Field South for transient and variable radio sources. This survey searches a deep ($<1$~mJy) part of time-domain parameter space on long timescales at 5.5~GHz. We found four sources that showed significant variability, for one of which the variability was at a very high level. No radio transients were detected and we place upper limits on the occurrence of such events. 

Through this survey we have explored the spectral and multi-wavelength properties of variables within the eCDFS. All of the variable sources had positive spectral indices and were associated with QSO and/or AGN. We conclude that the physical interpretation for one of the variables is best explained by episodic jet activity common to AGN. We conclude that three of the variables are young GPS sources whereby the variability and spectra are indicative of fairly recent activity within the radio jet. 

The radio spectra of these sources have implications for future variability surveys. Inverted spectra sources are quite rare within a typical radio survey and by definition, these sources will be more difficult to detect at lower frequencies (e.g. 1.4~GHz). Assuming that the variables discovered in this paper are drawn from a much larger population (over the whole sky), long duration variability surveys may be more efficient targeting higher frequencies to search for such objects.  

Papers describing variability surveys often quote the number of variables detected as a fraction/percentage of the total number of sources. This is misleading because it does not factor in the ability to detect different types and amounts of variability within those objects. Furthermore, the detection threshold is typically used as a proxy for the threshold at which variability could be detected. This is true, however a source at the detection threshold of a survey would have to have large and significant variability to be detected. Typically (at 5~GHz) we are dealing with fairly low levels of variability (from AGN activity) and this is only detectable in the brightest sample of sources. 

Within this dataset we have achieved sensitivities between 12.1 to 17.1~$\mu$Jy beam$^{-1}$ (1$\sigma$), utilising $\sim$250 hours of telescope time. We can clearly rule out extremely large amplitude variations in the bulk of sources. However, even in optimistic scenarios we are probably only capable of probing low levels of AGN type variability in sources $> \sim$~1~mJy. This is approximately two orders of magnitude higher than the sensitivity of this survey. It is also worth noting that the variables we have detected (which have inverted spectrum) constitute a fairly rare type of AGN ($\sim$10\% of the sample; \citealt{CDFS_paper}). The requirement to be both greater than $\sim$1~mJy (for variability to be detected) and belong to a class of object in the minority (i.e. GPS), means that we have detected variability in the majority of inverted spectrum sources that we were capable of detecting it within. We could therefore from a statistical standpoint conclude that variability is prevalent in these types of objects. 

Detections of transients still remain rare. Large area, sensitive, high cadence observations are now required to explore regions of parameter space not covered by previous surveys (e.g. see \citealt{metzger}). A number of next generation wide-field GHz radio interferometers that meet these requirements will soon survey the sky for transient and variable radio sources. These include: the Karoo Array Telescope (MeerKAT; \citealt{meerkat}) and the Australian Square Kilometre Array Pathfinder (ASKAP; \citealt{ASKAP}) and potentially the Square Kilometre Array (SKA; \citealt{SKA}). A comparative survey to this one over 1000~deg$^{2}$ with, for example, MeerKAT could detect may thousands of these variable sources. If an unknown transient population exists, leveraging the increased survey speeds of next generation instruments will be the optimum path to discovery.   
 
\section{Acknowledgements}
We thank the anonymous referees for their insightful comments which substantially helped to improve the paper. 
This work was supported by the Centre for All-sky Astrophysics (CAASTRO), an Australian Research Council Centre of Excellence (grant CE110001020) and through the Science Leveraging Fund of the New South Wales Department of Trade and Investment. The Australia Telescope Compact Array is part of the Australia Telescope National Facility which is funded by the Commonwealth of Australia for operation as a National Facility managed by CSIRO. DB and TM acknowledge funding through ARC grant DP110102034. This research has made use of NASAs Astrophysics Data System (ADS) and the NASA/IPAC Extragalactic Database (NED).

\appendix

\label{lastpage}

\end{document}